\documentclass[twocolumn,trackchange]{aastex62}

\graphicspath{{./}{figures/}}
\shorttitle{}
\shortauthors{Mondal \& Saha}

\begin{document}

\title{The AstroSat UV Deep Field North: The IRX - $\beta$ relation for the UV-selected galaxies at $z \sim 0.5 - 0.7$}

\author[0000-0003-4531-0945]{Chayan Mondal}
\affiliation{Inter-University Centre for Astronomy and Astrophysics, Ganeshkhind, Post Bag 4, Pune 411007, India}

\author[0000-0002-8768-9298]{Kanak Saha}
\affiliation{Inter-University Centre for Astronomy and Astrophysics, Ganeshkhind, Post Bag 4, Pune 411007, India}

\email{chayanm@iucaa.in, mondalchayan1991@gmail.com}

\begin{abstract}
The relation between the observed UV continuum slope ($\beta$) and the infrared excess (IRX) is used as a powerful probe to understand the nature of dust attenuation law in high-redshift galaxies. We present a study of 83 UV-selected galaxies between redshift 0.5 and 0.7 from the AstroSat UV Deep Field north (AUDFn) that encloses the GOODS-north field. Using empirical relation, we estimate the observed IRX of 52 galaxies that are detected in either one or both of the Herschel PACS 100$\mu$m and 160$\mu$m bands. We further utilize the multi-band photometric data in 14 – 18 filters from the UVIT, KPNO, HST, Spitzer, and Herschel telescopes to perform spectral energy distribution (SED) modeling. Both the observed and model-derived IRX - $\beta$ values show a large scatter within the span of previously known relations, signifying diversity in dust attenuation. We found a distinct relation between the best-fit power law slope of the modified Calzetti relation ($\delta$) in the IRX - $\beta$ plane: where the steeper SMC-like attenuation law prefers lower $\delta$ values. Our SED model based IRX - $\beta$ relation shows a preference for steeper SMC-like attenuation which we further confirm from the agreement between extinction-corrected star formation rates derived using H$\alpha$ emission line and the observed FUV plus reprocessed far-IR fluxes. The current study reveals a strong positive correlation between IRX and the galaxy stellar mass between $10^{9.5}$ and $10^{11.0}$ M$_{\odot}$, signifying increased dust production in more massive star-forming galaxies.

\end{abstract}

\section{Introduction}
\label{s_introduction}
Measuring the star formation rate (SFR) of galaxies from local to distant universe is one of the important goals of extra-galactic astrophysics. Among different SFR tracers, emission in the far-ultraviolet (FUV) band, which is primarily produced by young massive OB-type stars, is used as a direct probe to measure the recent SFR in a galaxy \citep{kennicutt1998,kennicutt2012}. But the presence of interstellar dust, which does not contribute majorly to the total baryonic mass of a galaxy, plays a crucial role by effectively attenuating the UV and optical photons through absorption or scattering \citep{calzetti2000,draine2007}. The energy associated with these lost FUV photons heats the dust which further emits in the far-infrared (FIR) band. The FIR emission, which scales with the dust column density, therefore can quantify the amount of dust-obscured star formation in a galaxy \citep{kennicutt1998,rieke2009}. Hence, deriving the total SFR of a galaxy would require the measurement of both the FUV and FIR continuum emission. However, this approach assumes minimal dust heating by the older stars which would bias the measurement of dust-obscured SFR using FIR emission \citep{buat2011,boquien2016}. Alternatively, modeling the multi-band spectral energy distribution (SED) of galaxies is considered a useful technique to derive total SFR including other galaxy parameters; such as stellar mass, star formation history, metallicity, stellar population age, etc, both for local \citep{kauffmann2003,salim2005} and high-redshift galaxies \citep{shapley2001,forster2004}. However, such energy-balanced SED modeling requires an accurate prescription of the dust attenuation law as a prior for robust estimation of these parameters including the FUV-based SFR. In other words, the knowledge of attenuation would help to break the age-dust degeneracy \citep{burgarella2005} while performing the SED modeling.

An attenuation law provides the value of effective extinction along the line of sight as a function of wavelength \citep{salim2020}. Owing to the proximity, attenuation laws are better studied in the case of the Milky Way and nearby galaxies. Several studies have empirically formulated extinction curves that show characteristic variation as a function of wavelength \citep{cardelli1989,calzetti1994,fitzpatrick1999,calzetti2001,gordon2003}. The two primary differences are the variation of slope in the extinction curve (significant in the shorter wavelength regime $\lambda<1700$~\AA) and the presence/absence of 2175~\AA~ UV bump \citep{stecher1965,gordon2003,noll2009}. Generally, the Milky Way-type dust extinction curve has a shallower slope including a UV bump \citep{fitzpatrick1999}, whereas the SMC-type curve is much steeper without the bump \citep{gordon2003}. Due to the difficulty in determining such laws at higher redshift, astronomers often assume one of these known attenuation laws to correct for dust extinction in distant galaxies, which might result in an incorrect measurement of dust-corrected quantities. Besides, the production and destruction of dust grains depend on different mechanisms which also vary with cosmic time as well as from one galaxy to another \citep{draine2009}. Specifically at higher redshift, galaxies undergo a different star-forming and dust production phase in a relatively metal-poor environment compared to the local galaxies \citep{marrone2018}. As a result, dust laws that apply to nearby galaxies might not work well for distant galaxies and an independent evaluation of attenuation law is desirable.

The idea of constructing empirical dust laws at higher redshift using FUV and FIR continuum emission was first shown by \citet{meurer1995} using observations of 9 nearby starburst galaxies. Their study showed that the UV continuum slope ($\beta$; which is defined as F$_{\lambda} \propto \lambda^{\beta}$, where F$_{\lambda}$ is the UV continuum flux), which indirectly traces the dust content, shows a clear relation with the infrared excess (IRX; which is defined as the ratio of FIR to FUV luminosity, i.e., L$_{FIR}$/L$_{FUV}$) \citep{calzetti1994,meurer1995,wilkins2013}. Later, \citet{meurer1999} confirmed the validity of a similar IRX - $\beta$ relation in $z\sim$3 U-dropout galaxies. For a set of starburst galaxies, \citet{gordon2000} showed a good agreement between attenuation of Balmer lines estimated using the IRX method and through independent analysis of radio observations, which reinforced the robustness of IRX - $\beta$ law. The IRX - $\beta$ relation soon became a promising technique for constraining dust attenuation at higher redshift, especially for galaxies without FIR observation \citep{bouwens2009}.

At higher redshift, several studies targeted various galaxy populations; for example, UV/optical-selected samples \citep{reddy2006,overzier2011,buat2012,marquez2016,reddy2018}, Lyman Break Galaxies (LBGs) \citep{heinis2014}, strongly lensed star-forming galaxies \citep{sklias2014}, to construct independent IRX - $\beta$ relation each signifying a specific attenuation law. \citet{hamed2023} studied 1049 star-forming galaxies at intermediate redshift $0.5 < z < 0.8$ (i.e., similar to the range targeted in our study) and reported a strong correlation of IRX - $\beta$ relation with metallicity, galaxy compactness, stellar mass, and specific SFR. The nature of IRX - $\beta$ is explored in many nearby star-forming galaxies as well, both at galactic \citep{takeuchi2012} and sub-galactic \citep{boquien2012,calzetti2021} scales. Several attempts are also made using analytical modeling \citep{ferrara2017} as well as galaxies from cosmological hydrodynamical simulations \citep{cullen2017,narayanan2018} to understand the dispersion in the empirical IRX - $\beta$ relations.

The IRX - $\beta$ relations, constructed for galaxies of different types and at different redshifts, show offset with respect to each other. Galaxies within the same sample also exhibit a large scatter about the IRX - $\beta$ curve. The scatter or the offset originates from multiple factors. \citet{kong2004} reported that the presence of older stellar populations with a wider age range can produce offset and scatter. Several other studies also associated stellar population age as the driver for the observed scatter \citep{burgarella2005,johnson2007,grasha2013}, whereas \citet{boquien2012} found that the scatter in the intrinsic UV continuum slope $\beta_{0}$ can produce dispersion in the observed IRX - $\beta$ values. Moreover, the location of a galaxy in the IRX - $\beta$ plane depends on dust grain size distribution, dust chemical compositions, star-dust geometry, dust column density, and radiative transfer of photons in the interstellar medium \citep{gordon1997,witt2000,inou2005,popping2017,narayanan2018}. \citet{salim2019} concluded that the varying slope of the attenuation curve, primarily driven by the changing optical opacity, is the main cause of scatter seen in the IRX - $\beta$ values.

At intermediate redshift $\sim$0.5, constraining the IRX - $\beta$ relation requires observation in the NUV band. \citet{hamed2023} used GALEX NUV observation that has a larger PSF size $\sim$5\farcs2, which increases the source blending issues resulting in non-reliable UV photometry. However, HST imaging, which has a superior angular resolution, could avoid this problem but only with limited sky coverage. The GOODS-N field, which we targeted in this study, is covered by two HST UVIS filters F275W and F336W but not with F225W (HDUV survey; \citet{oesch2018}). \citet{reddy2018} used the available HST UV imaging of galaxies in the GOODS-N field to constrain IRX -$\beta$ relation at redshift 1.5$<z<$2.5. Due to the unavailability of deep NUV imaging with larger sky coverage, the study of empirical IRX - $\beta$ relation at intermediate redshift is still limited. We used deep NUV imaging (angular resolution $\sim$1\farcs2) of the AstroSat UV Deep field north (AUDFn; \citet{mondal2023a}), observed using Ultra-Violet Imaging Telescope (UVIT) on AstroSat \citep{kumar2012,singh2014}, to improve the situation with more robust rest-frame FUV photometry of $z\sim$0.6 galaxies. We utilized the NUV source catalog of the GOODS-N field from \citet{mondal2023a} to study 83 galaxies that have a counterpart in the GOODS-N Herschel infrared source catalog between redshift 0.5 and 0.7 \citep{elbaz2011}. We acquired the observed $\beta$ measurements of these galaxies from \citet{mondal2023b} and estimated observed IRX using photometry from Herschel PACS 100$\mu$m and 160$\mu$m bands. We further performed SED modeling of the selected galaxies using photometry in 14 - 18 bands from FUV to Infrared and derived $\beta$ and IRX values. Our analyses highlight the nature of IRX - $\beta$ relation of UV-selected galaxies at a less explored redshift window. 

The paper is arranged as follows: Section \S\ref{s_data} presents the details of data and sample selection, Section \S\ref{s_analysis} describes the analysis, the results are discussed in Section \S\ref{s_results}, followed by a summary in \S\ref{s_summary}. Throughout the paper, we provide all magnitudes in the AB system, and adopt a cosmology with $H_0 = 70$ km s$^{-1}$ Mpc$^{-1}$, $\Omega_{\Lambda} = 0.7$, $\Omega_M = 0.3$\,.

\begin{figure}
    \centering
    \includegraphics[width=3.5in]{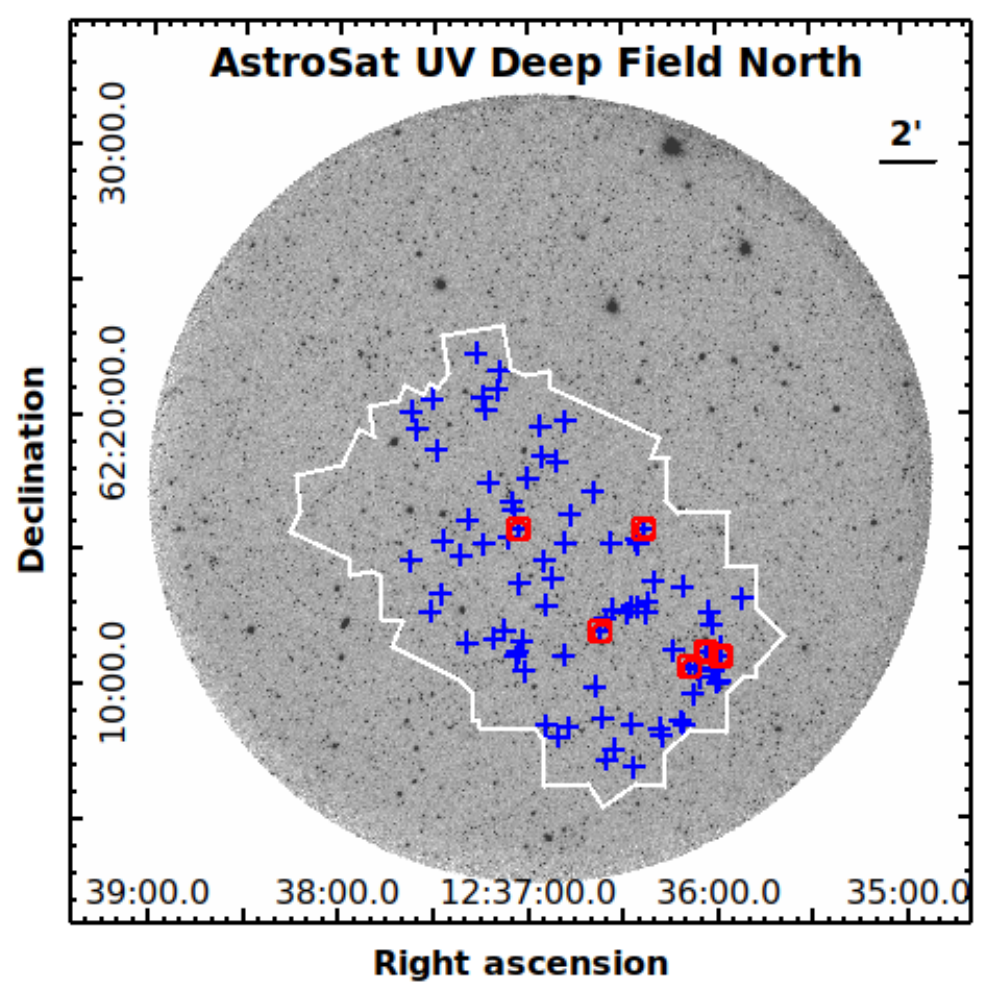} 
    \caption{The N242W band NUV image of the AstroSat UV Deep Field north (AUDFn). The white polygon shows the coverage by HST Cosmic Assembly Near-infrared Deep Extragalactic Legacy Survey (CANDELS; \citet{grogin2011,koekemoer2011}) of the GOODS-north deep field. The blue points mark the location of 83 galaxies selected in this study between redshift 0.5 and 0.7\,. We found 6 galaxies to be identified in the 2 Ms Chandra point-source catalog \citep{alexander2003} and marked them in red.}
    \label{fig_fov}
\end{figure}

\section{Data and Sample Selection}
\label{s_data}
To understand the relation between the UV continuum slope ($\beta$) and IRX, we essentially need observations in the rest-frame FUV, NUV, and FIR wavelengths. Our primary selection is based on the objects that are detected in the UVIT N242W band (see \citet{mondal2023a}) and have a reliable spectroscopic redshift measurement between 0.5 and 0.7 from the HST Grism (i.e., 3D-HST survey; \citet{momcheva2016}). The redshift range is fixed such that the UVIT N242W or N245M band could effectively sample the rest-frame FUV continuum of selected galaxies. Including the reliable redshift, we also checked the 3D-HST photometric catalog \citep{skelton2014} to select only those sources that do not have a neighboring source within a radius of 1\farcs4 i.e., the UVIT PSF FWHM. We found a total of 282 sources in the AUDFn catalog to satisfy these criteria (same as described in \citet{mondal2023b}). To identify a reliable FIR counterpart of these 282 objects, we considered sources listed in the Herschel GOODS-N catalog \citep{elbaz2011} and created a list by selecting only those with no neighbors within a radius of 4\farcs0 to minimize source confusion as specified in the Herschel data release manual. We cross-matched these Herschel sources with the 282 UV sources and found 83 common objects, which we considered as our final sample. To minimize the effect of source blending as well as background noise in the FIR band photometry, \citet{elbaz2011} used positional priors from the Spitzer 24$\mu$m image and performed PSF fitting to derive flux in the Herschel PACS bands (100$\mu$m and 160 $\mu$m). The detection in the Spitzer 24$\mu$m image was carried out using the priors from the Spitzer IRAC 3.6$\mu$m image obtained in the GOODS-Spitzer legacy program (PI - M. Dickinson). Finally, the sources only with $>3\sigma$ signal in each corresponding IR band are listed in the GOODS-N Herschel catalog \citep{elbaz2011}. The detection methodology adopted by \citet{elbaz2011} signifies that all of our final samples, though listed in the GOODS-N Herschel catalog, might not have detection in the PACS FIR bands. We found 53 out of the 83 UVIT-Herschel cross-matched samples to have measurements in the Herschel PACS bands (i.e., either in 100$\mu$m or in 160$\mu$m or both), which we could directly use to determine their FIR luminosity (as explained later in Section \S\ref{s_observed_beta_irx}). 

We acquired the UVIT photometry along with the estimated observed $\beta$ values of the final 83 UV-selected galaxies from our earlier study by \citet{mondal2023b}. We obtained their photometry in HST UVIS bands (F275W, F336W) from the HDUV catalog \citep{oesch2018}, HST ACS (F435W, F606W, F775W, F850LP), WFC3 IR (F125W, F140W, F160W), and Spitzer IRAC bands from the 3D-HST catalog \citep{skelton2014}, and the Spitzer MIPS 24$\mu$m and the Herschel PACS 100$\mu$m, 160$\mu$m bands from the Herschel GOODS-N catalog \citep{elbaz2011}. In Figure \ref{fig_fov}, we showed the location of the final samples in the AUDFn field. To identify potential AGNs in our samples, we utilized the 2 Ms X-ray source catalog of the GOODS-N field \citep{alexander2003} and found 6 sources (marked in red - Figure \ref{fig_fov}) that have X-ray counterparts. For deriving SFR using H$\alpha$ emission line flux, we used measurements from the HST CLEAR catalog \citep{simons2023} for a sub-sample of the selected galaxies.

\begin{figure}
    \centering
    \includegraphics[width=3.5in]{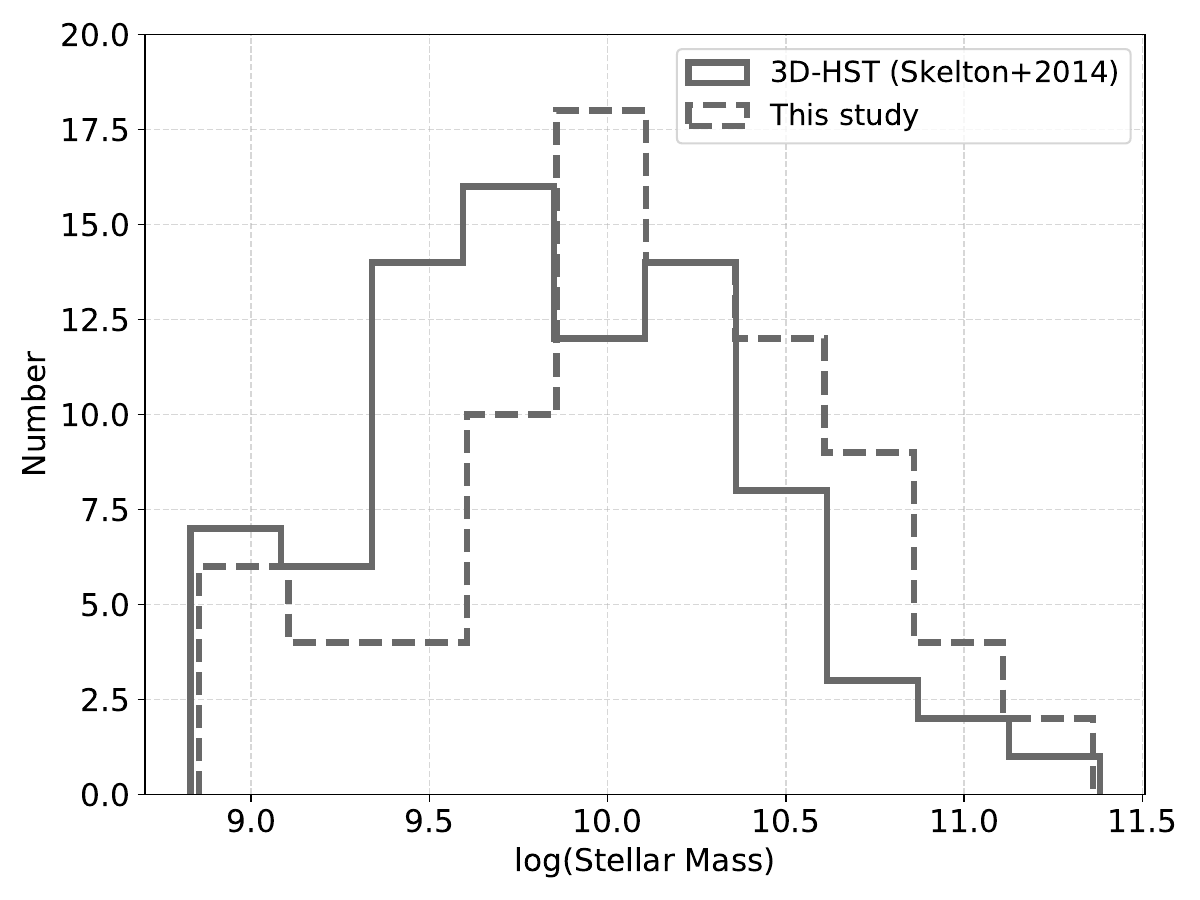} 
    \caption{Stellar mass distribution of the sample galaxies. The histogram in solid line is plotted using values from the 3D-HST catalog \citep{skelton2014}, whereas the dashed one shows measurements derived from SED modeling in this study.}
    \label{fig_mass_hist}
\end{figure}

In Figure \ref{fig_mass_hist}, we compare the stellar mass distribution of selected galaxies from the 3D-HST photometric catalog \citep{skelton2014} and as estimated through our SED analysis (discussed in Section \S\ref{s_model_beta_irx}) in this study. The stellar mass of our UV-selected galaxies has a range $\sim$ 10$^9$ - 10$^{11.5}$ M$_{\odot}$ with a peak around 10$^{10}$ M$_{\odot}$. We note here that our sample is incomplete due to different selection criteria. As discussed in \citet{mondal2023b}, we selected only clean UV sources (i.e., sources having no neighbor in the HST catalog within a radius of 1$\farcs$4) that have reliable spectroscopic redshift measurement from the HST G141 Grism observations. Both these conditions result in the exclusion of galaxies of different brightness, preferably on the fainter side. As the primary selection criterion demands detection in the UVIT NUV band, we missed galaxies that are not emitting enough in the rest-frame FUV and fall below the AUDFn 5$\sigma$ detection limit of 26.7 mag \citep{mondal2023a}.

\section{Analysis}
\label{s_analysis}
\subsection{Estimating observed IRX - $\beta$}
\label{s_observed_beta_irx}

\begin{figure}
    \centering
    \includegraphics[width=3.5in]{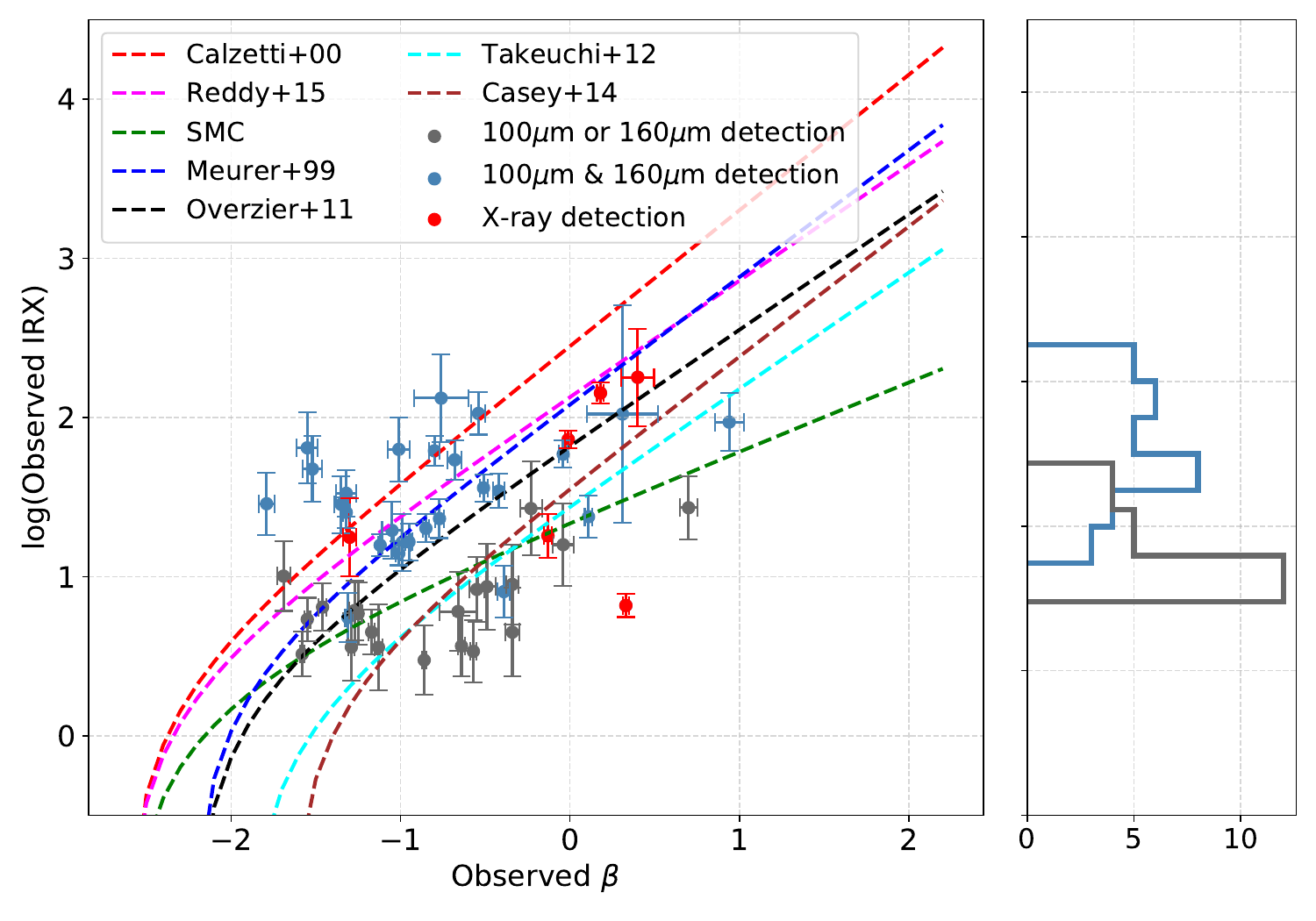} 
    \caption{The observed IRX (i.e., estimated using L$_{FIR}$ as derived from equation \ref{eq_irx}) and observed $\beta$ (i.e., acquired from \citet{mondal2023b}) of 52 galaxies among the total 83. 31 galaxies that are detected in both the PACS 100 $\mu$m and 160 $\mu$m bands are shown in blue, whereas the grey points represent 21 galaxies with detection in either of the two PACS bands. The IRX distribution of these two different classes is shown in adjacent histograms with the same respective color. Seven different IRX - $\beta$ relations, representing seven attenuation laws, are shown to convey the nature of IRX - $\beta$ applicable to our samples. Six of the total 83 sources (which also come within these 52 sources), that have a counterpart in the 2 Ms Chandra X-ray source catalog \citep{alexander2003}, are shown in red. The distribution signifies that our selected galaxies at redshift 0.5 - 0.7 have diversity in the attenuation law.}
    \label{fig_observed_irx_beta}
\end{figure}

As per the definition of IRX, the FIR luminosity (L$_{FIR}$) is estimated within the wavelength range 8 - 1000 $\mu$m \citep{gordon2000,burgarella2005}. The FUV luminosity (L$_{FUV}$) is defined at an effective wavelength of 1500~\AA. Hence, we utilized the M$_{1500}$ values reported in \citet{mondal2023b} for L$_{FUV}$. However, the defined FIR luminosity cannot be sampled using a single photometric band due to the huge wavelength range. \citet{helou1988} have formulated the total FIR luminosity in terms of measurements in 60 $\mu$m and 100 $\mu$m FIR bands. The relation is given in equation \ref{eq_irx}, where L(60$\mu$m) and L(100$\mu$m) are the luminosity of an object at its rest-frame wavelength of 60 $\mu$m and 100 $\mu$m, respectively. At the mean redshift of our sample (i.e., $z \sim 0.6$), the observed fluxes in the PACS 100 $\mu$m and 160 $\mu$m bands would respectively sample the 60 $\mu$m and 100 $\mu$m rest-frame fluxes and hence we used measurements in these two bands to calculate total observed FIR luminosity which is further used to estimate the observed IRX. We found 31 among the 83 selected galaxies to have detection in both the PACS 100 $\mu$m and 160 $\mu$m bands. Using equation \ref{eq_irx}, we estimated their IRX values which are shown as blue points in Figure \ref{fig_observed_irx_beta} along with their observed $\beta$ from \citet{mondal2023b}. 21 of our galaxies have detection in either of the two PACS bands. We estimated their IRX using the same equation and shown in the same figure in grey. Due to the non-detection in one PACS band, the observed IRX of these galaxies (i.e., grey points) would be smaller than the actual value. The adjacent IRX histograms shown in the same figure further demonstrate this. The error in the observed $\beta$, calculated from the photometric error of the fluxes, is obtained from \citet{mondal2023b}. We estimated the error in IRX from the photometric error in UVIT NUV, Herschel PACS 100 $\mu$m, and 160 $\mu$m bands that are used to measure the observed IRX. 

\begin{equation}
    L_{FIR} = 1.26(L_{60\mu m} + L_{100\mu m}),
    \label{eq_irx}
\end{equation}

To understand the nature of IRX - $\beta$ relation in this study, we have shown the trend of seven well-known relations from the literature (i.e., Meurer+99 \citep{meurer1999}, Calzetti+00 \citep{calzetti2000}, Reddy+15 \citep{reddy2015}, SMC type \citep{gordon2003}, Overzier+11 \citep{overzier2011}, Takeuchi+12 \citep{takeuchi2012}, and Casey+14 \citep{casey2014}) in Figure \ref{fig_observed_irx_beta}. These relations highlight the variation of attenuation in galaxies as discussed in Section \S\ref{s_introduction}. Galaxies having detection in both the PACS bands show a preference for a shallow extinction curve similar to Meurer+00, Calzetti+00, or Reddy+15. However, some of them, particularly with a redder UV continuum, support steeper SMC-type extinction law. We notice that 5 out of 6 potential AGNs have a relatively redder $\beta$ with a large range of IRX values.

\begin{figure}
    \centering
    \includegraphics[width=3.5in]{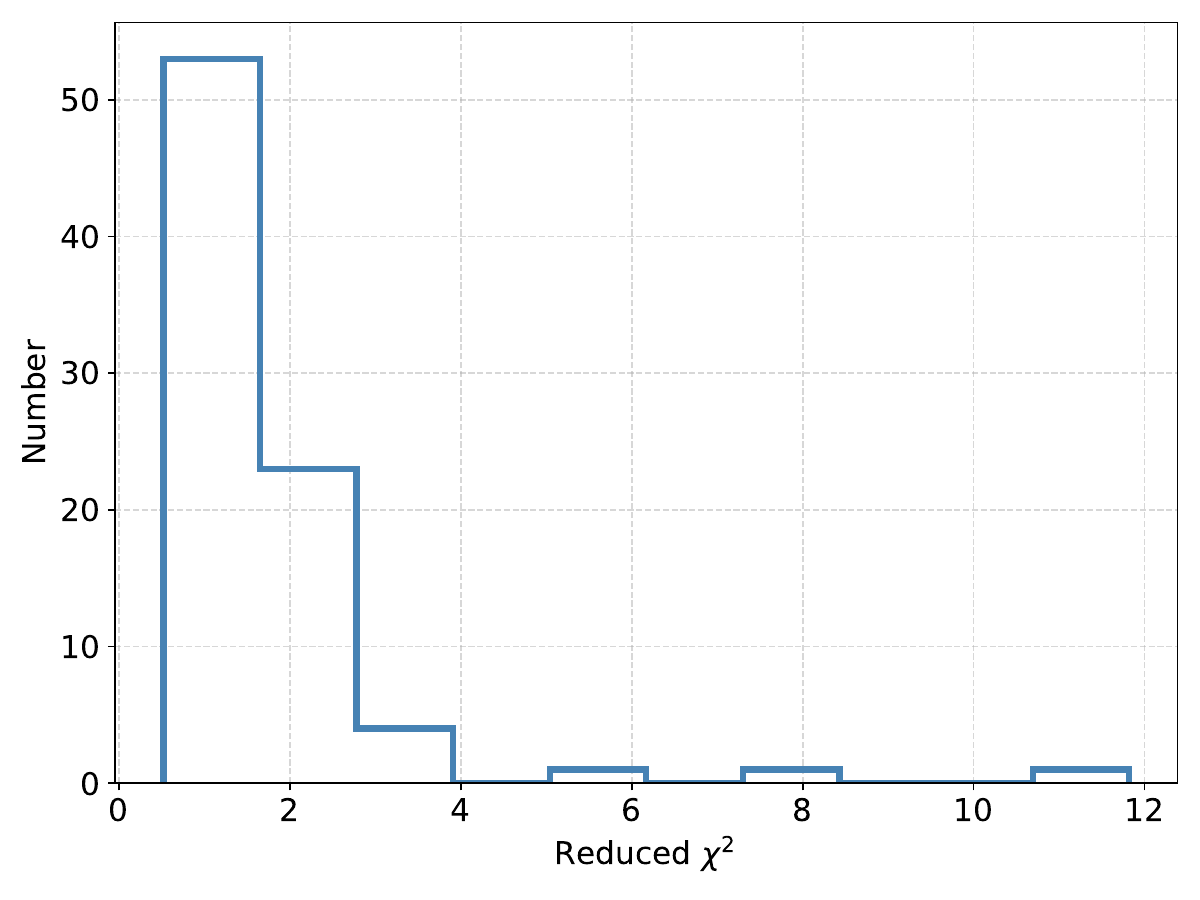} 
    \caption{The distribution of reduced $\chi^2$ values of best-fit CIGALE models for the selected galaxies. Only 5 among the total 83 galaxies has reduced $\chi^2$ higher than 3, which signifies the goodness of the SED fitting.}
    \label{fig_chi}
\end{figure}

\begin{table*}
\centering
\caption{Details of the photometric bands used in the CIGALE SED modeling}
\label{table_bands}
\begin{tabular}{p{3cm}p{3cm}p{2cm}p{2cm}p{3cm}}
\hline
Telescope & Filter & $\lambda_{mean}$ & Number of & Reference\\
 & & (~\AA~) & samples & \\
 (1) & (2) & (3) & (4) & (5)\\\hline
  AstroSat/UVIT & N242W & 2418 & 83 & \citet{mondal2023a}\\
   & N245M & 2447 & 83 & \\\hline
 HST WFC3/UVIS & F275W & 2710 & 32 & \citet{oesch2018}\\
 & F336W & 3355 & 32 & \\\hline
 KPNO & U & 3593 & 83 & \citet{skelton2014}\\\hline
 HST ACS & F435W & 4318 & 83 & \citet{skelton2014}\\
  & F606W & 6034 & 83 & \\
  & F775W & 7729 & 83 & \\
  & F850LP & 9082 & 83 & \\\hline
 HST WFC3/IR & F125W & 12516 & 83 & \citet{skelton2014}\\
 & F140W & 13969 & 82 & \\
 & F160W & 15396 & 83 & \\\hline
 Spitzer & IRAC1 & 35572 & 83 & \citet{skelton2014}\\
  & IRAC2 & 45049 & 83 & \\
  & IRAC3 & 57450 & 83 & \\
  & IRAC4 & 79158 & 82 & \\
  & MIPS 24$\mu$m & 24000 & 83 & \citet{elbaz2011}\\\hline
Herschel & PACS 100$\mu$m & 100000 & 41 & \citet{elbaz2011}\\
   & PACS 160$\mu$m & 160000 & 42 & \\
  
\hline
\end{tabular}

\textbf{Note.} Table columns: (1) name of the telescope; (2) name of the photometric band; (3) filter mean wavelength in \AA; (4) number of samples that have available photometric measurements in the band; (5) the source of the photometry.
\end{table*}

\subsection{Estimating IRX - $\beta$ through SED modeling}
\label{s_model_beta_irx}
We reproduced the measurements shown in Figure \ref{fig_observed_irx_beta} following a SED modeling approach for two primary reasons. First, 31 galaxies in the sample of 83 are not detected in either of the PACS 100 $\mu$m and 160 $\mu$m bands. As a result, we could not estimate their observed IRX values. To derive IRX for such sources, we solely depend on the SED modeling. Second, the value of observed FIR luminosity measured using equation \ref{eq_irx} assumes an approximation as it does not directly consider the entire wavelength range 8 - 1000 $\mu$m. Therefore, the observed IRX may differ from the actual value based on the measurement of FIR luminosity alone. Ideally, the value of L$_{FIR}$ should be estimated by integrating the spectrum of each galaxy between 8 and 1000 $\mu$m, which requires SED modeling. Here, we considered all 83 galaxies and their available multi-band photometric measurements to perform SED modeling using the CIGALE \citep{boquien2019}. The details of the photometric bands including the number of identified galaxies in each of them are listed in Table \ref{table_bands}. 

\begin{table*}
\caption{The values of CIGALE input parameters used in the SED modeling}
\label{table_cigale}
\begin{tabular}{p{4cm}p{4.5cm}p{8.5cm}}
\hline
Parameters &  & Values\\\hline

& Star formation history - sfh2exp & \\\hline
Age & & 500, 1000, 2000, 3000, 4000, 5000, 6000, 7000, 8400 (Myr)\\
$\tau_{main}$ & & 500, 1000, 2000, 4000, 6000 (Myr)\\
Burst$_{age}$ & & 250 (Myr)\\
$\tau_{burst}$ & & 10, 50 (Myr)\\
$f_{burst}$ & & 0.02, 0.05, 0.1\\\hline

 & Stellar population model - BC03 \citep{bruzual2003} & \\\hline
 Stellar IMF & & Chabrier \citep{chabrier2003}\\
 Metallicity & & 0.004, 0.008, 0.02\\\hline

 & Nebular & \\\hline
logU &  & $-$3.0\\
Gas metallicity & & 0.001\\
Electron density & & 100\\
f\_esc & & 0.0\\\hline
 & Dust  attenuation law - Modified Calzetti 2000 & \\\hline

E(B$-$V) & & 0.0, 0.05, 0.1, 0.15, 0.2, 0.3, 0.4, 0.5, 0.6, 0.8, 1.0, 1.2, 1.5, 1.8\\
E(B$-$V) factor & & 0.44\\
UV bump wavelength & & 2175~\AA \\
UV bump width & & 350~\AA \\
UV bump amplitude & & 0 \\
Power law slope ($\delta$) & & -0.7,-0.5,-0.3,-0.1, 0.0, 0.1, 0.3\\
qpah & & 2.5\\
U$_{min}$ & & 10.0\\
$\alpha$ & & 2.0\\
$\gamma$ & & 0.02\\
\hline
\end{tabular}
\end{table*}

\begin{figure*}
\centering
    \includegraphics[width=7in]{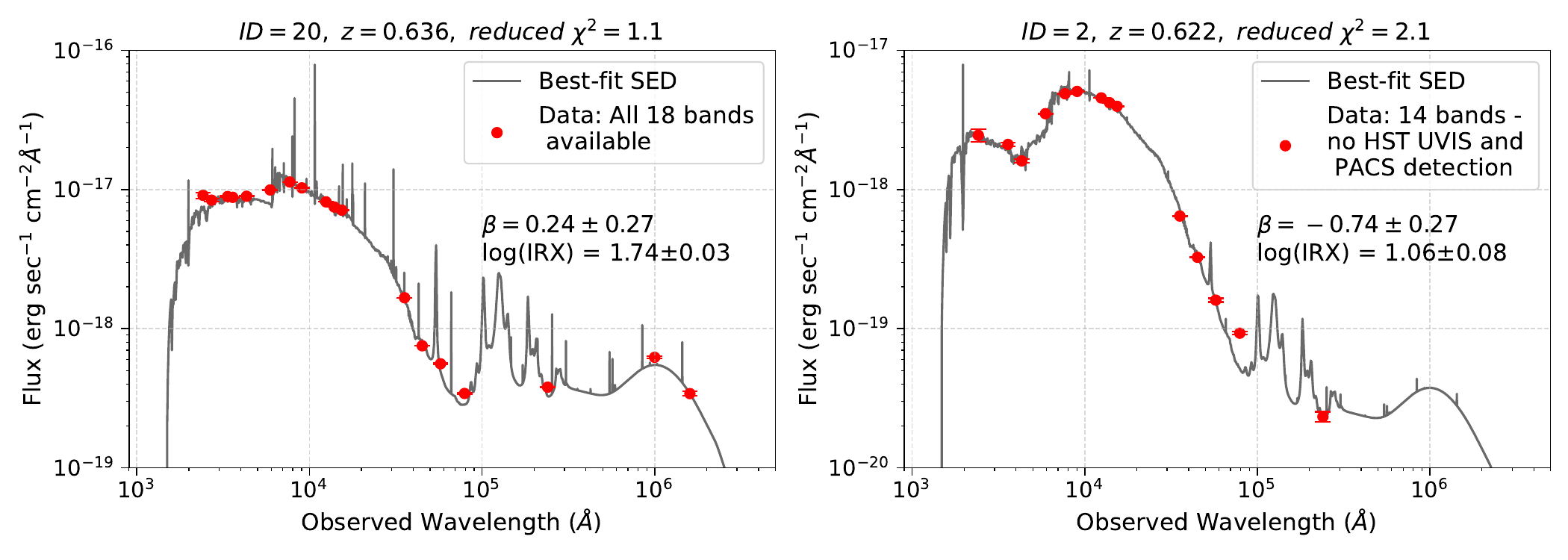}\\
    \includegraphics[width=7in]{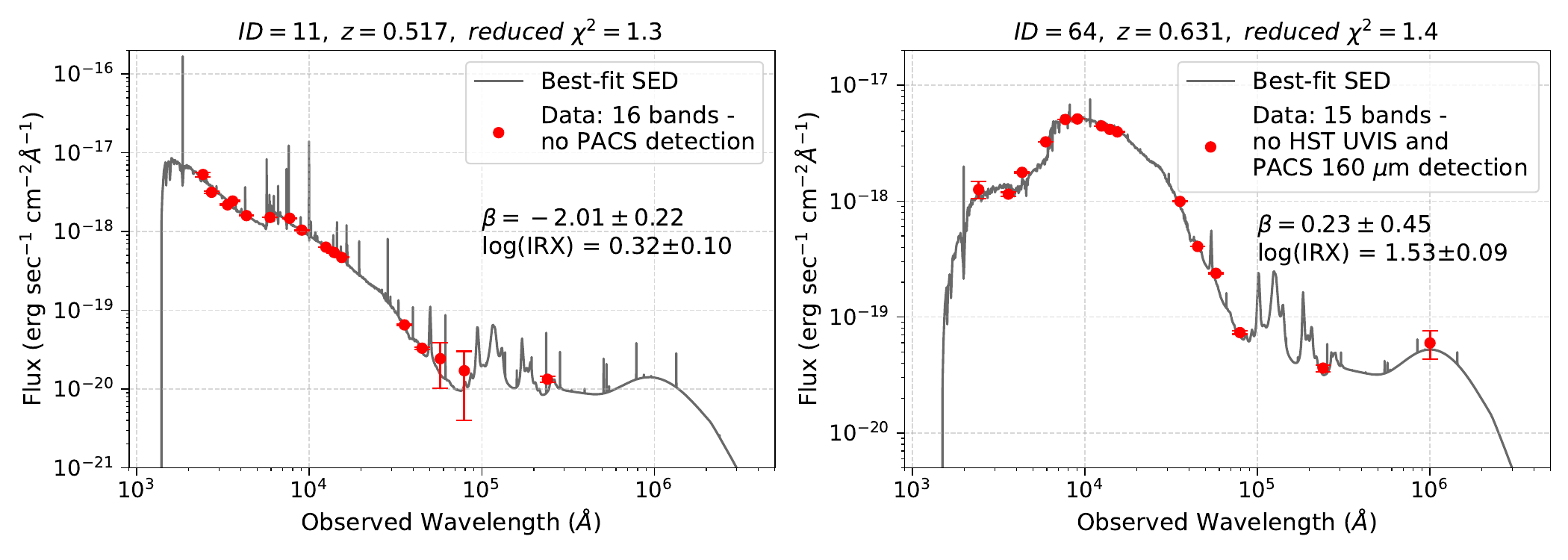}
    \caption{The best-fit SEDs (grey solid line) along with the observed fluxes (red points) of four selected galaxies. The sample ID, redshift, reduced $\chi^2$, model-derived observed $\beta$ and IRX values are noted in each figure. Each galaxy shown here presents a distinct case in terms of the availability of photometric data (ID 20: all 18 bands available; ID 2: 14 bands - no HST UVIS and PACS detection; ID 11: 16 bands - no PACS detection; ID 64: 15 bands - no HST UVIS and PACS 160 $\mu$m detection.}
    \label{fig_sed}
\end{figure*}

Some of our selected galaxies do not have measurements in the HST UVIS bands as they fall outside the observing coverage; whereas the unavailability of the Herschel PACS photometry means their non-detection in those FIR bands. We have adopted an approach to utilize the maximum number of available bands among those listed in Table \ref{table_bands} to perform the SED modeling for each galaxy. All the flux values acquired from different catalogs (as described in \S\ref{s_data}) are corrected for the Milky Way Galactic extinction before feeding to the CIGALE. We used the Schlegel \citep{schlegel1998} Galactic extinction map housed in the \textit{NASA/IPAC Infrared Science Archive} \footnote{https://irsa.ipac.caltech.edu/applications/DUST/} for the V band Galactic extinction (A$_V$). The Galactic extinction in each of the listed bands is then calculated assuming the Fitzpatrick \citep{fitzpatrick1999} extinction law ($R_{V}$ = 3.1). 

To derive the best-fit model for each galaxy, we used a CIGALE input configuration as specified in Table \ref{table_cigale}. We chose an exponentially decaying star formation history (SFH) combined with a recent starburst event for the model SEDs. The e-folding time of the main SFH is varied between 500 and 6000 Myr to account for SFRs with different decay rates. As each of the galaxies shows rest-frame FUV emission, we added a recent starburst of age 250 Myr with e-folding times of 10 and 50 Myr and different burst strengths (i.e., $f_{burst}$) to be incorporated in the SFH models. We used the BC03 stellar population models \citep{bruzual2003} of three different metallicities (i.e., Z = 0.004, 0.008, 0.02) with the Chabrier initial mass function \citep{chabrier2003}. The UV-selected star-forming galaxies at a particular redshift are expected to show less variation in their intrinsic UV continuum slope \citep{leitherer1999,calzetti2001}. Hence the parameter space, chosen for SFH and metallicity in CIGALE, offers enough scope to pick up any significant variation in the intrinsic dust-free UV slope. 

As our primary goal is to understand the nature of dust attenuation law, we have given more freedom to the dust parameters. We adopted the \textit{dustatt\_modified\_starburst} module of CIGALE which considers modification in the actual Calzetti attenuation law $k_{C}(\lambda)$ \citep{calzetti2000}. The modified law ($k(\lambda)$), as given in equation \ref{eq_dustlaw}, incorporates the 2175~\AA~ UV bump ($D_{bump}(\lambda)$; \citep{stecher1965,noll2009}) and also allows variation in the slope of the attenuation curve by a parameter $\delta$. The allowed variation of the slope in Modified Calzetti law spans other known attenuation curves e.g., \citet{fitzpatrick1999,cardelli1989,odonnell1994,gordon2003,reddy2015} that can be aped in the SED modeling of the selected galaxies. We selected 14 values for the reddening E(B$-$V) in the range 0.0 - 1.8 and 7 different values of $\delta$ between $-0.7 - 0.3$. The values of the other dust parameters are fixed as noted in Table \ref{table_cigale}. We have fixed the parameters related to the 2175~\AA~ UV bump as the broadband filters used in the modeling are not sensitive enough to pick up the bump-strength variation. As pointed out by \citet{salim2019}, the variation in the attenuation law is primarily driven by the slope of the extinction law, whereas the variation of bump strength plays a minimal role. Hence, the choice of our dust parameters is optimal for understanding the variation in attenuation.

\begin{figure}
    \centering
    \includegraphics[width=3.5in]{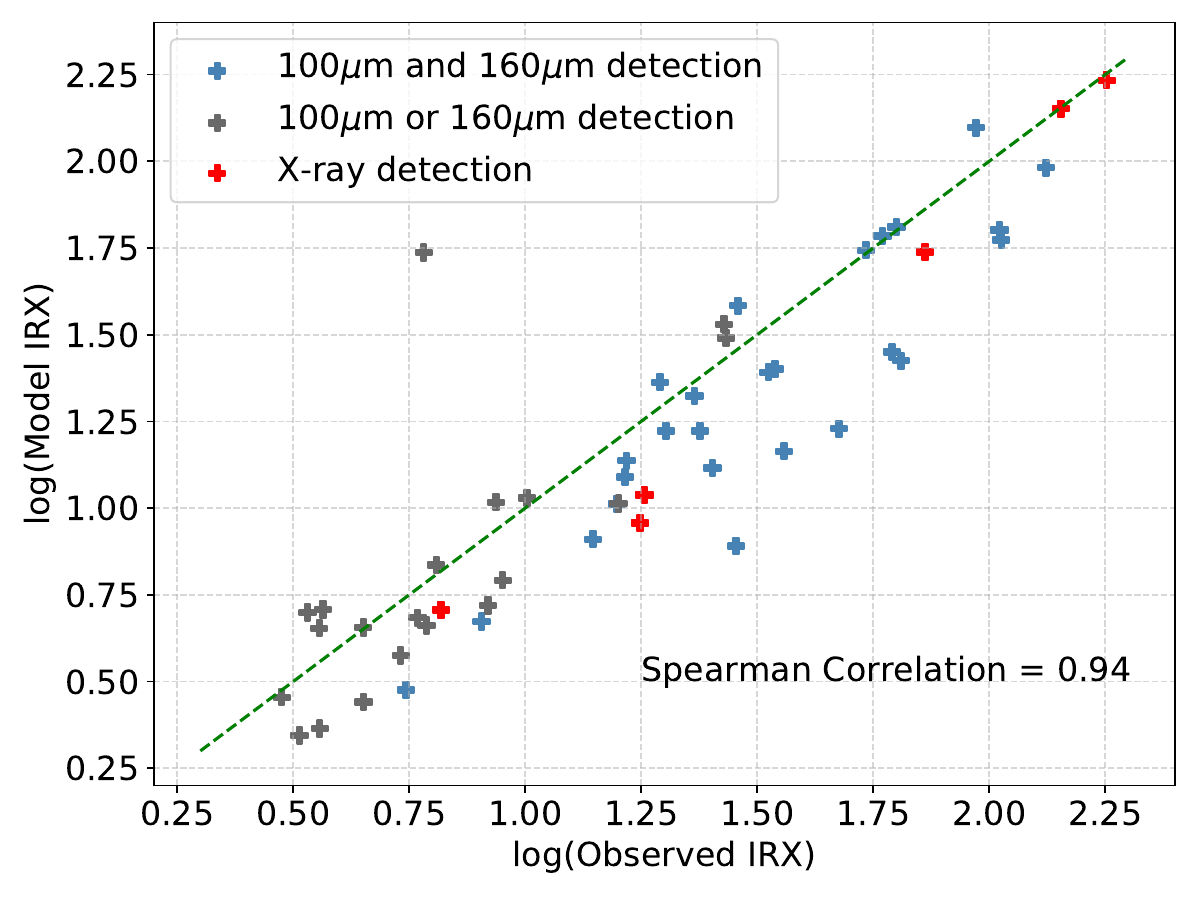} 
    \caption{The comparison between observed and model-derived IRX values for 53 galaxies, which have PACS FIR detection (either in 100$\mu$m or 160$\mu$m or both bands). The green dashed line shows the 1:1 correspondence. The overall good correlation signifies the robustness of the model-derived IRX. Six samples with X-ray counterparts in the 2 Ms Chandra catalog are marked in red.}
    \label{fig_irx_comp}
\end{figure}

\begin{equation}
    k(\lambda) = D_{bump}(\lambda) + \left[k_{C}(\lambda)\times (\frac{\lambda}{\lambda_{V}})^{\delta} \right],
    \label{eq_dustlaw}
\end{equation}

\begin{figure*}
    \centering
    \includegraphics[width=7in]{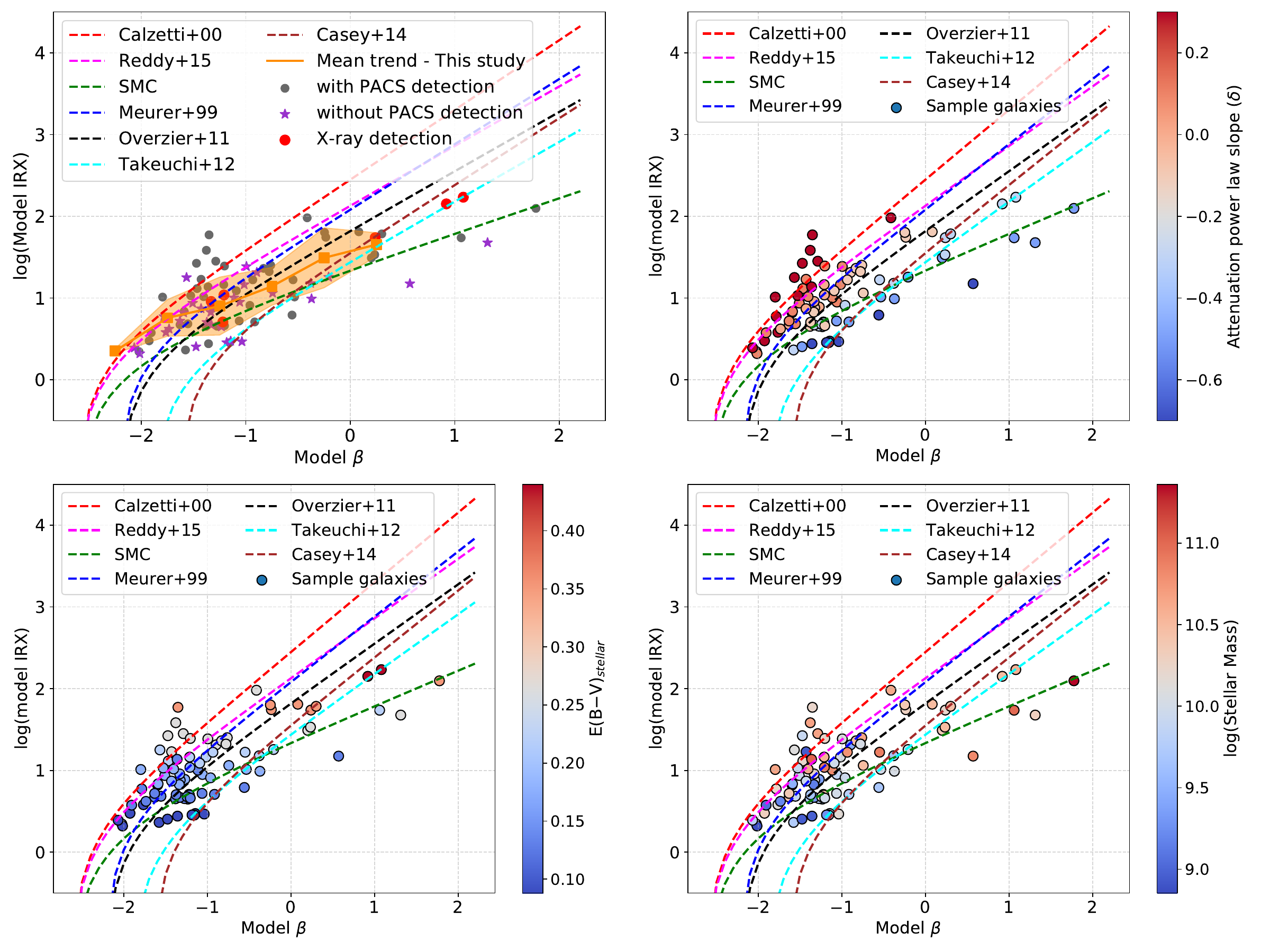} 
    \caption{The values of IRX and $\beta$ of 83 galaxies derived from the CIGALE SED modeling. Seven different IRX - $\beta$ relations, representing seven attenuation laws, are shown to convey the nature of attenuation applicable to our samples. \textit{Top-left:} The grey points denote galaxies that have observed IRX measurement as discussed in Section \S\ref{s_observed_beta_irx}. Galaxies without PACS detection (i.e., without observed IRX measurements) are shown by purple stars. Six samples with an X-ray counterpart in the 2 Ms Chandra catalog are marked in red. The orange points are the mean IRX estimated in equal-width $\beta$ bins of size $\Delta \beta$ = 0.5. The orange solid line shows the mean model-derived IRX - $\beta$ trend in our study. The orange shaded area denotes 1$\sigma$ scatter in IRX. The distribution shows a similar trend as noticed in Figure \ref{fig_observed_irx_beta}, signifying the robustness of the SED modeling in this study. The overall trend of the mean IRX with $\beta$ indicates an SMC-like attenuation curve. \textit{Top-right}: The color bar shows the value of power law slope $\delta$ as described in equation \ref{eq_dustlaw}. The model SED supporting a more negative value of $\delta$ favors steeper SMC-type law, whereas galaxies with higher values of $\delta$ follow greyer attenuation laws. \textit{Bottom-left:} The color bar shows the value of internal reddening of the stellar continuum (where, E(B$-$V)$_{stellar}$ = 0.44$\times$E(B$-$V)). The figure shows that with an increasing value of E(B$-$V) galaxies move up along the IRX - $\beta$ relation. \textit{Bottom-right:} The color bar shows the value of total stellar mass derived from SED modeling in this study. The majority of the low mass galaxies ($\log(M_{*})~<$ 10.0) have lower IRX values (i.e., log(IRX) $<$ 1.0), whereas the massive galaxies ($\log(M_{*})>$ 10.0) mostly show larger FIR emission with log(IRX) $>$ 1.0.}
    \label{fig_model_irx_beta_collage}
\end{figure*}

The SED fitting is performed in batch mode. The best-fit model for each galaxy is determined using the $\chi^2$ minimization technique as explained in \citet{boquien2019}. In Figure \ref{fig_chi}, we showed the distribution of reduced $\chi^2$ values of all the galaxies. We found 82\% of the galaxies to have reduced $\chi^2$ less than 2.0 which highlights the robustness of the fitting. Following the fitting, we derived the best-fit values of each parameter including the best-fit model SED. As specific outputs, CIGALE also provides the observed UV continuum slope $\beta$ and the IRX based on the respective best-fit SED for each galaxy. CIGALE derives observed $\beta$ by fitting a straight line in log(F$_{\lambda}$) - log($\lambda$) plane of the best-fit template within wavelength 1268 - 2580 \AA~ as specified by \citet{calzetti1994}. We showed fitted SEDs of four galaxies chosen to have photometry available in 14 (lowest) to 18 (highest) wavebands in Figure \ref{fig_sed} to highlight the fitting in the presence/absence of HST UVIS and/or Herschel PACS data. Especially, the two galaxies (ID = 2 and 11), without PACS detection, show the importance of SED modeling to constrain the FIR continuum compared to cases that have PACS FIR photometry (ID = 20 and 64). To validate the robustness of the SED modeling, we show the observed (from \S\ref{s_observed_beta_irx}) and the model-derived IRX of all the galaxies with PACS detection in Figure \ref{fig_irx_comp}. The results show good agreement between both the quantities (spearman correlation coefficient = 0.94) highlighting consistency in our modeling. This correlation further certifies to trust the model-derived IRX of 30 galaxies which are not detected in the PACS FIR bands and hence could not have an observed IRX measurement. In Figure \ref{fig_model_irx_beta_collage}, we have shown the $\beta$ and IRX of all 83 galaxies estimated from the SED modeling. We notice a distribution similar to the observed quantities shown in Figure \ref{fig_observed_irx_beta}. The galaxies without an FIR PACS measurement also fall within the range of the other galaxies in the IRX - $\beta$ plane. Here, we found a clearer trend of redder galaxies following the SMC-type attenuation law. We note here that a larger allowed variation of the input parameters, that has been fixed to a single value (for example, the nebular parameters), produces an insignificant change in the best-fit values of IRX, $\beta$, and stellar mass.

\section{Results}
\label{s_results}

This study explores the nature of attenuation law in galaxies between redshift 0.5 and 0.7 through IRX - $\beta$ relation. Our results, shown in Figure \ref{fig_observed_irx_beta} and Figure \ref{fig_model_irx_beta_collage}, highlight the diverse dust properties in galaxies at this epoch with the reddest galaxies more clearly favoring SMC-type extinction. Among different model parameters (Table \ref{table_cigale}), we found $\delta$ and E(B$-$V) to show a clearer trend in the IRX - $\beta$ relation, whereas galaxy stellar mass shows a positive correlation with the IRX.

\subsection{The trend of IRX - $\beta$ relation}
\label{s_trend}

We investigate the average trend in our IRX - $\beta$ relation to find out the most favorable attenuation law. In Figure \ref{fig_model_irx_beta_collage} (top-left), we considered equal-size bins in $\beta$ of width $\Delta \beta$ = 0.5 and estimated the mean and standard deviation in IRX for galaxies within each bin. We have excluded six samples that have an X-ray counterpart while deriving these mean values. The orange points and the shaded region shown in the figure respectively denote the measured mean IRX and corresponding 1$\sigma$ scatter. The overall trend is more similar to the SMC-like IRX - $\beta$ relation which signifies a steeper extinction curve in these galaxies. As we have sample incompleteness arising due to different selection criteria (discussed in \S\ref{s_data}), the observed trend in IRX - $\beta$ does not represent the nature of entire galaxy populations at this redshift. However, the majority of the galaxies with a redder UV continuum ($\beta > -0.5$) preferably follow SMC-like dust law, whereas those with a relatively bluer continuum ($\beta < -0.5$) show a diverse nature of attenuation laws.

The overall distribution of our sample galaxies in Figure \ref{fig_observed_irx_beta} and Figure \ref{fig_model_irx_beta_collage} lies within the range of known IRX - $\beta$ relations including a significant scatter. Such variation signifies different attenuation laws which indicate changes in dust grain properties or the star-dust geometry across the studied galaxies \citep{witt2000,gordon1997,inou2005,narayanan2018}. \citet{salim2019} reported significant scatter in the IRX - $\beta$ values of 23000 low-redshift galaxies at 0.01~$<~z~<~0.30$ which originates due to the diverse nature of attenuation driven by variable slope and 2175~\AA~bump strength in the extinction laws. Similar dispersion in the observed IRX - $\beta$ is also reported in the higher redshift $z\sim3$ LBGs by \citet{marquez2019}. \citet{boquien2009,boquien2012} reported such scatter at the sub-galactic scales of nearby normal star-forming and starburst galaxies. We notice $\sim$ 1.5 dex variation in IRX values within $\beta$ $-1$ to $-2$ where the majority of our galaxies lie. This plausibly signifies notable variation in the dust size distribution or star-dust orientation in these galaxies at $z\sim0.6$.

\begin{figure}
    \centering
    \includegraphics[width=3.5in]{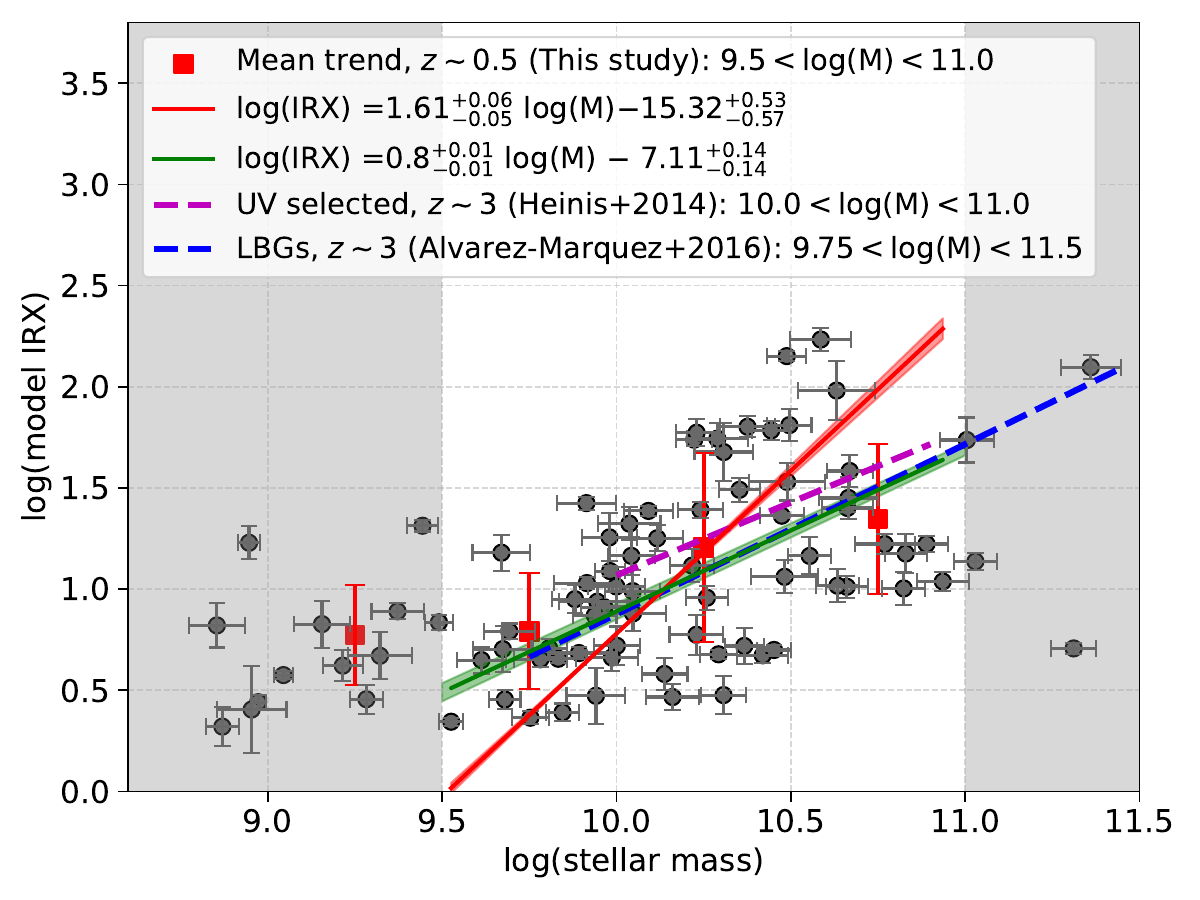}
    \caption{IRX and stellar mass of 83 sample galaxies derived from the CIGALE SED modeling (grey points). The red points show the mean IRX value estimated for four equal-width mass bins of size $\Delta~\log(M_{*})$ = 0.5 between $\log(M_{*})$ = 9.0 and $\log(M_{*})$ = 11.0\,. The red solid line, derived from an MCMC resampling considering error in both log(IRX) and $\log(M_{*})$, shows the relation of log(IRX) - log(Stellar Mass) in the mass range 9.5$<\log(M_{*})<$11.0. The green solid line, derived considering error only in log(IRX), agrees well with the relation derived for UV-selected and Lyman Break galaxies at $z\sim3$ by \citet{marquez2016,heinis2014}. The 1$\sigma$ confidence range of both the fitted relations (red and green lines) are shown by shaded regions in respective colors.}
    \label{fig_irx_mass}
\end{figure}

\subsection{Relation with $\delta$ and E(B$-$V)}
\label{s_delta}

Among the different dust parameters listed in Table \ref{table_cigale}, we allowed the power law slope $\delta$ and E(B$-$V) to vary over a large range for constraining the nature of dust laws through SED modeling. E(B$-$V) traces the amount of dust present in the galaxy, whereas the deviation of the attenuation curve from the well-known Calzetti law is modeled using the $\delta$ parameter (as shown in equation \ref{eq_dustlaw}; \citet{noll2009,boquien2019}). The parameters related to the UV bump, mass fraction of PAH (qpah), minimum radiation field (U$_{min}$), power law slope of the variable radiation field ($\alpha$), dust mass fraction linked to young star-forming regions ($\gamma$) are fixed to values suitable for star-forming galaxies. For each galaxy, CIGALE provides the best-fit value of $\delta$ and E(B$-$V) including many other parameters. In Figure \ref{fig_model_irx_beta_collage} (top-right), we have shown the model-derived $\beta$ and IRX values of all the galaxies including a color axis representing the best-fit $\delta$ values. Our results show a clear relation of $\delta$ with varying IRX - $\beta$ relations. The galaxies that follow the SMC-like dust have more negative $\delta$ values (i.e., extinction curve steeper than the Calzetti curve), whereas $\delta$ gradually shifts to higher values for galaxies supporting greyer extinction curves like Meurer+99, Calzetti+00, or Reddy+15. Following an equivalent modeling approach using CIGALE, \citet{buat2012} reported similar behavior of $\delta$ for 751 UV-selected galaxies at $0.95 < z < 2.2$. This relation portrays the steepness of different attenuation curves that apply to individual galaxies based on their location in the IRX - $\beta$ plane. For a fixed value of intrinsic UV continuum slope ($\beta_{0}$), a smaller value of $\delta$ would signify a steeper extinction law \citep{boquien2012}. The origin of such dependency lies in the star-dust geometry where a complex clumpy dust distribution moves a galaxy vertically upward in the IRX - $\beta$ plane favoring a shallower extinction curve \citep{gordon1997,witt2000,narayanan2018}. Whereas galaxies with a simpler star-dust geometry would move the galaxy downward in the IRX - $\beta$ plane.

As the derived value of $\delta$ is sensitive to intrinsic UV continuum slope ($\beta_{0}$), it is important to know the order of diversity in $\beta_{0}$ which denotes the diverse SFH of the selected galaxies. Around 87\% of our sample galaxies have $\beta_{0}$ within a small range between $\sim$ $-2.4$ and $-1.8$. We found only $\sim$6\% of the galaxies to have observed $\beta$ within the same above range which signifies the diversity of dust is much more effective than the intrinsic SFH in our sample. We also noticed no specific correlation in the IRX - $\beta$ relation with stellar population age and metallicity within the range of variation allowed in the SED modeling. Therefore, the relation of $\delta$ in the IRX - $\beta$ plane noticed in this study reflects the diversity in the attenuation laws at redshift $\sim$ 0.6\,. 

Variation in the effective optical depth has an important contribution to the total attenuation \citep{salim2019,salim2020}. Considering differences in the scattering of bluer and redder photons for the lower and higher optical depth limits, \citet{chevallard2013} used radiative transfer models to show that galaxies having an optically thinner dusty medium would follow a steeper SMC-like extinction curve. In contrast, higher optical depth would make the extinction curve shallower. This effect is seen in Figure \ref{fig_model_irx_beta_collage} (bottom-left) which shows that the increase in E(B$-$V) (i.e., an increase in the optical depth which can be due to higher dust column density) moves the galaxy upward in the IRX - $\beta$ plane. 

\subsection{Relation with stellar mass}
\label{s_mass}
As the amount of dust content in galaxies is often related to their total stellar mass, we explored such a relation in our study. We used the model-derived best-fit stellar mass values to color code the same IRX - $\beta$ plot and show it in Figure \ref{fig_model_irx_beta_collage} (bottom-right). The results signify no strong trend between the total stellar mass and the type of attenuation law. However, we noticed that the reddest galaxies ($\beta > 0$), favoring the SMC-like dust law, are mostly massive ($\log(M_{*})>$ 10.5). 

The infrared excess is known to have a strong correlation with galaxy stellar mass, where massive galaxies show more dust emission \citep{reddy2010,whitaker2014,whitaker2012b,marquez2016}. This correlation is reported to show no significant evolution with redshift \citep{whitaker2014,pannella2015}, although \citet{whitaker2012b} found a larger scatter in derived IRX values. \citet{buat2012,heinis2014} already showed such a correlation for UV-selected galaxies. Considering error in both IRX and stellar mass, we performed Markov chain Monte Carlo (MCMC) resampling of a linear relation of the form log(IRX)~=~m$\times \log(M_{*})$~+~c, where m and c are the slope and intercept of the proposed linear relation between log(IRX) and $\log(M_{*})$, within the mass range 9.5$<\log(M_{*})<11$. We performed 1000 iterations with priors ranging 1~$<$~m~$<$~2 and $-20<$~c~$<-10$ and obtained the best-fit values of m and c as $1.61^{+0.06}_{-0.05}$ and $-15.32^{+0.53}_{-0.57}$, respectively. The relation derived from this exercise is quoted in equation \ref{eq_irx_mass} (i.e., red line in Figure \ref{fig_irx_mass}). A similar MCMC resampling of the linear relation considering error only in the IRX results in a slightly different solution as shown by the green line in the same figure. Concerning the relations reported in the literature, we notice an overall similar trend between the derived IRX and the stellar mass for our UV-selected galaxies (Figure \ref{fig_irx_mass}). The galaxies with relatively low mass ($\log(M_{*})<$ 10.0) tend to have lower IRX values (log(IRX) $<$ 1) which signify low dust column density. However, we have a larger sample incompleteness in the low-mass regime as the heavily dust-obscured low-mass star-forming galaxies might be missing if they are not detected in the rest-frame FUV but have a higher IRX value. On the higher mass side ($\log(M_{*})>$ 10), where we have better sample completeness, we noticed the galaxies mostly have higher IRX values (log(IRX) $>$ 1). We have also shown the trend of mean IRX estimated at four different stellar mass bins, each having more than 5 galaxies, between $\log(M_{*})$ 9.0 and 11.0 in the same figure.
Our MCMC-derived empirical relation between log(IRX) and $\log(M_{*})$ (i.e., equation \ref{eq_irx_mass}) agrees well with relations found by \citet{marquez2016} and \citet{heinis2014} respectively for Lyman Break and UV-selected galaxies at $z\sim3$. This signifies that irrespective of the nature of attenuation, the massive galaxies have produced more dust during their evolution.

\begin{equation}
\log(IRX) = 1.61^{+0.06}_{-0.05}~\times~\log(M_{*}) - 15.32^{+0.53}_{-0.57}
\label{eq_irx_mass}
\end{equation}

\begin{figure*}
    \centering
    \includegraphics[width=7in]{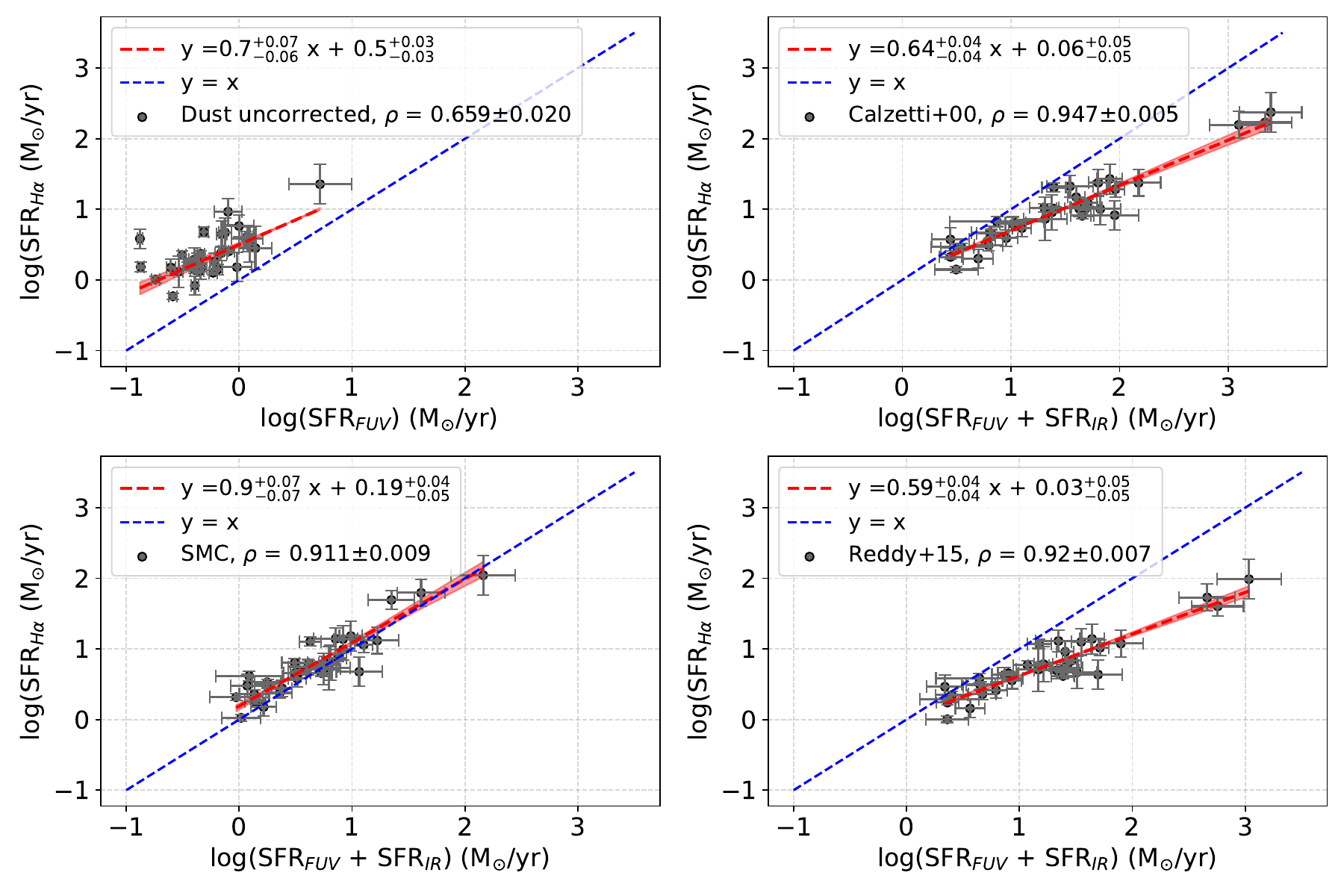} 
    \caption{The correlation between SFR derived using different tracers is shown. \textit{Top-left}: SFR derived from H$\alpha$ and FUV fluxes without galaxy's internal extinction correction. The other three figures show SFR after correcting for extinction assuming Calzetti+00 \textit{(top-right)}, SMC \textit{(bottom-left)}, and Reddy+14 \textit{(bottom-right)} dust laws. The SFR$_{FUV}$ + SFR$_{FIR}$ signifies the total SFR derived from observed FUV flux plus the IRX-derived reprocessed FIR flux. The red line in each figure shows the best-fit solution of the linear relation (of the form $y = m\times x + c$) derived through an MCMC resampling using 1000 iterations and priors ranging 0~$<$~slope~(m)~$<$~2 and $-2~<$~intercept~(c)~$<~3$. The red shaded region marks the 1$\sigma$ confidence range of each fitted line. The best-fit relation for each is noted in respective panels. The blue dashed line in each plot shows 1:1 line, whereas the quoted $\rho$-value denotes the median Pearson correlation coefficient including 1$\sigma$ difference estimated from 1000 iterations considering the error in respective SFR values}. The dust-corrected SFRs show the best agreement for the SMC-type attenuation law.
    \label{fig_clear_sfr}
\end{figure*}

\section{Testing the validity of SMC-like attenuation law}

\label{s_sfr}
The average trend of model IRX - $\beta$ relation, as shown by the orange line in Figure \ref{fig_model_irx_beta_collage} (top-left), prefers an SMC-like behavior. However, such a trend cannot be considered as the generalized one in the targeted redshift regime due to the scatter as well as the sample incompleteness we have in our study. Here, we attempt to find the nature of attenuation law using a different approach. We used the recently released HST CLEAR catalog of the GOODS-N field \citep{simons2023} that offers spectroscopic redshift of a large number of galaxies. We note here that the redshift provided in the CLEAR catalog has an insignificant difference from the 3D-HST redshift (which was used to derive observed $\beta$ in \citet{mondal2023b}) with $\Delta~z$/(1+$z$) 0.0002 $\pm$ 0.0001. This implies a mean redshift difference of 0.00012 which is extremely small to create any significant effect in the derived $\beta$ values. The HST CLEAR catalog also provides fluxes of detected emission lines for each source. We cross-matched our 83 selected galaxies with the CLEAR catalog and found 38 common objects. The CLEAR survey is carried out using the WFC3 IR G102 grism which covers the H$\alpha$ line of galaxies between redshift 0.5 and 0.7 in the observed frame. We found 35 out of 38 cross-matched galaxies to show H$\alpha$ emission with SNR $> 3$ and considered those for further analysis.
\begin{equation}
    SFR_{H\alpha} (M_{\odot}~yr^{-1}) = 5.37 \times 10^{-42} L_{H\alpha} (erg~sec^{-1})
    \label{eq_sfr_halpha}
\end{equation}
\begin{equation}
    SFR_{FUV} (M_{\odot}~yr^{-1}) = 4.42 \times 10^{-44} L_{FUV} (erg~sec^{-1})
    \label{eq_sfr_fuv}
\end{equation}
\begin{equation}
    SFR_{FIR} (M_{\odot}~yr^{-1}) = 3.88 \times 10^{-44} L_{FIR} (erg~sec^{-1})
    \label{eq_sfr_fir}
\end{equation}

We estimated star formation rates of these 35 galaxies using both the $H\alpha$ and FUV fluxes after correcting for the MW Galactic extinction as explained in \S\ref{s_model_beta_irx}. The H$\alpha$ based SFR (SFR$_{H\alpha}$) is calculated using the relation from \citet{murphy2011} as given in equation \ref{eq_sfr_halpha}, where L$_{H\alpha}$ is the luminosity in erg~sec$^{-1}$. Similarly, the rest-frame FUV luminosity (L$_{FUV}$), derived from the UVIT NUV data, is used to calculate the FUV-based SFR (SFR$_{FUV}$) of the galaxies using equation \ref{eq_sfr_fuv} adopted from \citet{murphy2011}. Both the SFRs, derived without internal dust correction, are shown in the top-left panel of Figure \ref{fig_clear_sfr}. The SFR$_{H\alpha}$ consistently shows a higher value than SFR$_{FUV}$ for all the galaxies. As the FUV emission traces star formation in the last $\sim$ 100 Myr contrary to the H$\alpha$ emission which is sensitive only up to 10 Myr, this result plausibly indicates a recent starburst in these galaxies. The other possibility is the significant effect of dust attenuation which can produce this mismatch in the derived SFRs if not corrected. As the FUV photons are most sensitive to the presence of dust, the deficit in the SFR$_{FUV}$ indicates a substantial amount of dust-obscured star formation in the UV-selected sample galaxies. Therefore, an accurate estimation of SFR, derived using either FUV or H$\alpha$ line emission, requires internal dust correction. As the dust content in galaxies evolves with cosmic time \citep{kurcz2014,finkelstein2012}, the correction for internal extinction is also crucial for knowing the actual evolution of SFR density with redshift \citep{pillepich2018,zavala2021}.

To check the correlation of dust-corrected SFR$_{H\alpha}$ and SFR$_{FUV}$, we assumed three different extinction laws (i.e., Calzetti+00, SMC, Reddy+15) and estimated the dust-obscured SFR for each. Using the observed $\beta$ and the FUV luminosity (from \citet{mondal2023b}) in the respective IRX - $\beta$ relation, we estimated FIR luminosity and further used that to derive dust-obscured SFR$_{FIR}$ following equation \ref{eq_sfr_fir} from \citet{murphy2011}. We plotted the total SFR (i.e., SFR$_{Tot}$ = SFR$_{FUV}$ + SFR$_{FIR}$) along with the internal dust-corrected SFR$_{H\alpha}$ in the top-right, bottom-left, and bottom-right of Figure \ref{fig_clear_sfr} assuming Calzetti+00, SMC, and Reddy+15 type extinction laws, respectively. The internal extinction correction for the SFR$_{H\alpha}$ is also performed using respective A$_V$ calculated from corresponding $\beta$ - E(B$-$V) relations. As shown in Figure \ref{fig_clear_sfr}, the value of the pearson correlation coefficient has improved from 0.659$\pm$0.020 to 0.947$\pm$0.005, 0.911$\pm$0.009, and 0.92$\pm$0.007 after considering dust-correction as per Calzetti+00, SMC, and Reddy+15 type extinction laws, respectively. Moreover, among these three, the best agreement between SFR$_{Tot}$ and SFR$_{H\alpha}$ is noticed for the SMC-type dust. This plausibly signifies that our sample galaxies most preferably follow SMC-type attenuation law which was also noticed in the model-derived IRX - $\beta$ relation.

\section{Summary}
\label{s_summary}
We used multi-band photometric observations to study the nature of attenuation law in 83 UV-selected galaxies between redshift 0.5 and 0.7 in the GOODS-north field. We acquired the measurements of UV continuum slope ($\beta$) and FUV luminosity from \citet{mondal2023b}, whereas to estimate FIR luminosity we utilized the Herschel GOODS-N catalog provided by \citet{elbaz2011}. We further performed SED modeling using observations in 18 bands from FUV to FIR and derived $\beta$, IRX, stellar mass, E(B$-$V), slope of the modified Calzetti law, etc. The main findings of this work are summarized below:

\begin{itemize}
    \item We estimated $\beta$ and IRX values of our sample galaxies directly from observed data as well as using CIGALE SED modeling. The selected galaxies show a large scatter both in the observed and model-derived IRX - $\beta$ plane. Such scatter signifies diverse dust properties that require variation in the attenuation laws.

    \item The model-derived IRX values show $\sim$ 1.5 dex variation between $\beta$ $-2$ and $0$ which contains $\sim$ 87\% of the samples.

    \item We noticed a preference for SMC-like attenuation law in the mean IRX - $\beta$ trend as derived from the SED modeling.

    \item The power law slope $\delta$, which indicates deviation from Calzetti dust law, shows a systematic trend in the IRX - $\beta$ relation. The galaxies preferring shallower attenuation laws (e.g., Meurer+99, Cazetti+00, Reddy+18) have progressively higher $\delta$ values, whereas SMC-like steeper attenuation goes along with smaller $\delta$ values.

    \item We found that an increase in the E(B$-$V), which signifies an increase in optical depth, shifts galaxies upward in the IRX - $\beta$ plane.

    \item We found $\sim$ 87\% of our galaxies to have intrinsic UV continuum slope ($\beta_{0}$) within a smaller range from $-2.4$ to $-1.8$ signifying less diversity in the SFH and further reinforming the variation of dust attenuation as the primary driver behind the observed scatter in the IRX - $\beta$ values.

    \item We do not see a clearer trend in the IRX - $\beta$ relation with respect to galaxy mass. However, the redder galaxies with $\beta > 0$ are mostly massive ($\log(M_{*})>$ 10.5) and they favor SMC-like attenuation law.

    \item We noticed a positive correlation between derived IRX and galaxy stellar mass (equation \ref{eq_irx_mass} for the range 9.5~$<~\log(M_{*})~<$~11.0) which signifies more efficient dust production in massive galaxies.

    \item The dust-corrected SFRs, estimated using H$\alpha$ emission line (SFR$_{H\alpha}$) and the observed FUV added with reprocessed FIR fluxes (SFR$_{FUV}$ + SFR$_{FIR}$), show best agreement with each other while assuming SMC-like extinction law. This signifies a preference for steeper attenuation law among the selected galaxies at redshift $\sim$ 0.6.

\end{itemize}

\section*{Acknowledgements}

We thank the anonymous referee for valuable suggestions. This work is primarily based on observations taken by AstroSat/UVIT. UVIT project is a result of collaboration between IIA, Bengaluru, IUCAA, Pune, TIFR, Mumbai, several centers of ISRO, and the Canadian Space Agency. This work also uses observations taken by the 3D-HST Treasury Program (HST-GO-12177 and HST-GO-12328) with the NASA/ESA Hubble Space Telescope, which is operated by the Association of Universities for Research in Astronomy, Inc., under
NASA contract NAS5-26555. This research made use of Matplotlib \citep{matplotlib2007}, Astropy \citep{astropy2013,astropy2018}, community-developed core Python packages for Astronomy, and SAOImageDS9 \citep{joye2003}.

\software{SAOImageDS9 \citep{joye2003}, Matplotlib \citep{matplotlib2007}, Astropy \citep{astropy2013,astropy2018}}


\end{document}